\documentclass[twocolumn,aps,prc,showpacs,superscriptaddress,preprintnumbers,floatfix,nofootinbib]{revtex4}
\usepackage{epsfig,graphics}
\usepackage{graphicx}
\usepackage{dcolumn}
\usepackage{bm}
\usepackage{amsmath}
\usepackage[usenames]{color}
\usepackage{ulem} 

\usepackage{CJK}

\begin{document}
\begin{CJK*}{GBK}{song}

\title{Forward-backward elliptic anisotropy  correlation in parton cascade}

\author{ L. X. Han}
\affiliation{Shanghai Institute of Applied Physics, Chinese Academy of Sciences, P.O. Box 800-204, Shanghai
201800, China} \affiliation{Graduate School of the Chinese Academy of Sciences, Beijing 100080, China}
\author{ G. L. Ma}
\affiliation{Shanghai Institute of Applied Physics, Chinese Academy of Sciences, P.O. Box 800-204, Shanghai
201800, China}
\author{ Y. G. Ma}
\thanks{Corresponding author: Email: ygma@sinap.ac.cn}
\affiliation{Shanghai Institute of Applied Physics, Chinese Academy of Sciences, P.O. Box 800-204, Shanghai
201800, China}
\author{ X. Z. Cai}
\affiliation{Shanghai Institute of Applied Physics, Chinese Academy of Sciences, P.O. Box 800-204, Shanghai
201800, China}
\author{ J. H. Chen}
\affiliation{Shanghai Institute of Applied Physics, Chinese
Academy of Sciences, P.O. Box 800-204, Shanghai 201800, China}
\author{ S. Zhang}
\affiliation{Shanghai Institute of Applied Physics, Chinese
Academy of Sciences, P.O. Box 800-204, Shanghai 201800,
China}
\author{ C. Zhong} \affiliation{Shanghai Institute of
Applied Physics, Chinese Academy of Sciences, P.O. Box 800-204,
Shanghai 201800, China}

\date{ \today}

\begin{abstract}
A potential experimental probe, forward-backward elliptic
anisotropy correlation ($C_{FB} $), has been proposed by Liao and
Koch to distinguish the jet and true elliptic flow contribution to
the measured elliptic flow ($v_2$) in relativistic heavy-ion
collisions. Jet and flow fluctuation contribution to elliptic flow
is investigated within the framework of a multi-phase transport
model using the $C_{FB} $ probe. We found that the $C_{FB} $
correlation is remarkably different and is about two times of that
proposed by Liao and Koch. It originates from the correlation
between fluctuation of forward and backward elliptic flow at low
transverse momentum, which is mainly due to the initial
correlation between fluctuation of forward and backward
eccentricity. This results in an amendment of the $C_{FB}$  by a
term related to the correlation between fluctuation of forward and
backward elliptic flow. Our results suggest that a suitable
rapidity gap for $C_{FB} $ correlation studies should be around
$\pm$ 3.5.
\end{abstract}


\maketitle

The results from Brookhaven Relativistic Heavy-Ion Collider (RHIC)
indicate that a strongly-interacting partonic matter has been
created in relativistic nucleus-nucleus
collisions~\cite{RHIC_white_paper}. Two powerful probes exposing
the characteristics of the new matter are elliptic flow and jet.
Elliptic flow, has been measured via the second Fourier
coefficient ($v_2$) in the azimuthal distribution of final
particles~\cite{Voloshin, Jinhui}. The $v_2$ data show remarkable
hydrodynamical behaviors, which implies the formed matter is
thermalized in a very short time and expands collectively as a
liquid with low shear viscosity/entropy. On the other hand, jet,
which is produced non-collectively by initial hard scatterings,
has been experimentally studied by nuclear modification factor and
jet-like
correlation~\cite{5prl,GLV,Phenix10,AbelevPLB,AdamsPRL,AbelevPRL,
song_guoliang}. These observation shows that jet losses energy
when it passes through the hot and dense QCD medium formed in
heavy-ion collisions. However, two probes, elliptic flow and jet,
are correlated in fact. The jet contributes to the anisotropy in
final azimuthal distribution and certainly to the observed $v_2$,
especially for high transverse momentum ($p_T$) because of the
path length dependence of energy loss in non-central
collisions~\cite{Gyulassy,Axel}. Nuclear community usually
classify such a non-collective behavior as non-flow. Therefore, it
is important for elliptic flow measurement to distinguish these
two parts from observed $v_2$ to further separate collective and
non-collective properties of the new matter. A potential
experimental probe, forward-backward elliptic anisotropy
correlation ($C_{FB} $), has been proposed for this purpose by
Liao and Koch (LK) ~\cite{Liao}. It is believed that the new probe
can distinguish how jet and elliptic flow contribute to the
observed $v_2$ which n probably reveal the momentum scale where
elliptic flow or (semi-) hard processes dominate. Using a
two-component parameterization, they found the F-B correlation
($C_{FB} $), which takes a maximum value of unity at low
transverse momentum ($p_T$) and falls to zero with $p_T$. This
indicates the jet contribution to the measured $v_2$ grows with
increasing $p_T$ and even jet contribution dominates at high
$p_T$. However, event-by-event fluctuation of $v_2$ is also
important for $v_2$ measurement~\cite{Axel,fluctuation}.
Experimental data shows that the relative nonstatistical
fluctuations of the $v_2$ is to be approximately
40$\%$~\cite{fluctuation}. Unfortunately it is not taken into
account in LK's calculation. Our present study considers the
effect of $v_2$ fluctuation to further understand bulk properties
and related jet effects in the early stage of heavy-ion
collisions. Our paper presents the relative nonstatistical
fluctuation effects on the forward-backward $v_2$ correlation
($C_{FB} $) and the $G$-factor range which reflects jet
contribution to observed $v_2$ with a multi-phase transport model
(AMPT)~\cite{AMPT}.  A more generalized $C_{FB}$ which includes
F-B elliptic flow fluctuation correlation is formulated.

The AMPT model consists of four main components: the initial
conditions, partonic interactions, conversion from partonic to
hadronic matter and hadronic interactions. The initial conditions,
which include the spatial and momentum distributions of minijet
partons and soft string excitations, are obtained from the Heavy
Ion Jet Interaction Generator model. Scatterings among partons are
modeled by Zhang's Parton Cascade model, which at present includes
only two-body scatterings in which  cross sections are obtained
from the pQCD calculation by screening mass. In the version with
string melting mechanism, partons include minijet partons and
partons from melted strings. The quark coalescence model is used
to converte partons into hadrons. The dynamics of the subsequent
hadronic matter is then described by a relativistic transport
model. The detail of the AMPT model can be found in
Ref.~\cite{AMPT}. From the previous AMPT calculations, it is found
that  elliptic flow can be built by strong parton
cascade~\cite{AMPT,SAMPT,AMPT_v2_rap,AMPT_charm_v2,AMPT_sys_v2,Jinhui}
and jet losses energy into partonic medium to excite a Mach-like
cone structure~\cite{amptdihadron,amptgammajet,ampthotspots}. It
is clear from the above that partonic effect can not be neglected.
Therefore the string
melting AMPT version is 
appropriate when the energy density is much higher than the
critical density which is predicted for phase transition. In this
work, we use the string melting AMPT model to simulate Au+Au
collisions at $\sqrt{s_{NN}}$ = 200 GeV. The partonic interaction
cross section is set to 10 mb.

Elliptic flow $v_2$ is defined as
     \begin{equation}
      <v_2(p_T)> \,= \frac{\int_{0}^{2\pi}d\phi\cos2\phi<\frac{d^2N}{p_tdp_td\phi}>}
      {\int_0^{2\pi} d\phi <\frac{d^2N}{p_tdp_td\phi}>}
      \equiv \frac{<V_2(p_t)>}{<\frac{dN}{p_{t}dp_t}>},
      \end{equation}
where the total yields $\frac{d^2N}{p_t dp_td\phi}$ are resulted
from both elliptic flow and jet~\cite{Liao} which then contribute
to the measured $v_2$.

The proposed observable $C_{FB}[p_T]$ is the correlation of the
total elliptic flow $V_2[p_T]$ between forward ($F$) and backward
($B$) rapidity bins~\cite{Liao}, defined as:
      \begin{equation}
      C_{FB}[p_T] = \frac{ \langle V_2^FV_2^B \rangle}{\langle V_2^F\rangle \langle V_2^B\rangle}.
      \label{eq2}
      \end{equation}

    With the assumption $ \langle  \eta^F \rangle  \,= \,\langle \eta^B\rangle \,= 0$, \,$\langle \xi\rangle \,= \,\langle 1-\xi\rangle \,= \frac{1}{2}$ and $\xi(1-\xi)= 0$, $C_{FB}$ changes to
        \begin{align}
         C_{FB} = \,
     &
     \frac{(1-g)^2\langle v_2^f\rangle ^2+2g(1-g)\langle v_2^f\rangle \langle v_2^j\rangle }{[(1-g)\langle v_2^f\rangle +g\langle v_2^j\rangle ]^2}
     \nonumber
     \\
  &
  +\frac{4\langle \eta^F\eta^B\rangle (1-g)^2\langle v_2^f\rangle ^2}{[(1-g)\langle v_2^f\rangle +g\langle v_2^j\rangle ]^2},
\label{eq3}
\end{align}
where $\eta^{F(B)}$ represents random deviations from the average
elliptic flow yield in forward (backward) rapidity bin,
$v_2^{f(j)}$ represents elliptic flow (jet) part in $v_2$, $\xi$
represents the jet contribution to $F$ or $B$ rapidity bin, and
$g$ factor, giving the relative weight of the jet contribution to
the total, i.e. (F+B), yield, is defined as~\cite{Liao}
      \begin{equation}
      g = \frac{\int_{0}^{2\pi}d\phi\langle \frac{dN^j}{d\phi}\rangle }
      {\int_{0}^{2\pi}d\phi\langle \frac{dN^f}{d\phi}\rangle +\int_{0}^{2\pi}d\phi\langle \frac{dN^j}{d\phi}\rangle }.
      \label{eq4}
      \end{equation}
When the correlation of the fluctuation of F-B elliptic flow yield
is zero, (i.e. $\langle \eta^F\eta^B\rangle  \,= 0$), then LK
found:
      \begin{equation}
      C_{FB} =
      \frac{(1-g)^2\langle v_2^f\rangle ^2+2g(1-g)\langle v_2^f\rangle \langle v_2^j\rangle }{[(1-g)\langle v_2^f\rangle +g\langle v_2^j\rangle ]^2}.
      \label{eq5}
      \end{equation}
When only jet contributes to $v_2$, that is $g = 1$, $C_{FB} = 0$;
On the other hand, if only elliptic flow contributes to $v_2$,
then $g = 0$, $C_{FB} = 1$.

However, if the correlation of the fluctuation of F-B elliptic
flow  is nonzero and for example $\langle \eta^F\eta^B\rangle  \,=
\frac{1}{4}$, $C_{FB} $ will change to
      \begin{equation}
      C_{FB} =
      \frac{2(1-g)^2\langle v_2^f\rangle ^2+2g(1-g)\langle v_2^f\rangle \langle v_2^j\rangle }{[(1-g)\langle v_2^f\rangle +g\langle v_2^j\rangle ]^2}.
      \label{eq6}
      \end{equation}
When only jet contributes to $v_2$, then $g = 1$, $C_{FB} = 0$; However, if only elliptic flow contributes to $v_2$, then $g = 0$, $C_{FB} = 2$.

We introduce a $G$ factor to represent total jet contribution
$V_2^j$ to total $V_2(p_T)$, which has a relation to $g$ factor in
\cite{Liao} as followed:

        \begin{align}
      G =  \frac{\langle v_2^j\rangle \int_{0}^{2\pi}d\phi\langle \frac{dN^j}{d\phi}\rangle }
{\langle v_2\rangle (\int_{0}^{2\pi}d\phi\langle
\frac{dN^f}{d\phi}\rangle +\int_{0}^{2\pi}d\phi\langle
\frac{dN^j}{d\phi}\rangle )} = \frac{v_2^j}{v_2}g.
        \label{eq7}
        \end{align}

Using a different form of total $V_2(p_T)$ in forward ($F$) and
backward ($B$) rapidity bins, i.e.
      $V_2^F = (\frac{1}{2}+\eta^F)(1-G)\langle V_2\rangle +\xi G\langle V_2\rangle
      $ and
      $V_2^B = (\frac{1}{2}+\eta^B)(1-G)\langle V_2\rangle +(1-\xi) G\langle V_2\rangle $,
  we get the formula below with the assumption $\langle
\eta^F\rangle  \,= \,\langle \eta^B\rangle  \,= \, 0$, $\langle
\xi\rangle  \,= \,\langle 1-\xi\rangle \,= \frac{1}{2}$, and
$\xi(1-\xi) = 0$, but $\langle \eta^F\eta^B\rangle \,\neq 0$:
      \begin{equation}
      C_{FB} =
      1 - G^2 + 4(1-G)^2
      \langle \eta^F\eta^B\rangle .
      \label{eq10}
      \end{equation}
When the correlation of the fluctuation of F-B elliptic flow is
zero, (i.e. $\langle \eta^F\eta^B\rangle  \,= 0$),
      $C_{FB} = 1 - G^2 $. From here,  only jet will contribute to $v_2$ when $G =
1$, $C_{FB} = 0$. On the other hand,   only elliptic flow
contributes to $v_2$ when $G = 0$, $C_{FB} = 1$. While when the
correlation is nonzero, for example $\langle \eta^F\eta^B\rangle =
\frac{1}{4}$,
      $C_{FB} = 2(1 - G)$.
 Then we can find that  only jet contributes to $v_2$ with a
 condition
$G = 1$, $C_{FB} = 0$. However,  only elliptic flow will
contribute to $v_2$ when $G = 0$, $C_{FB} = 2$.

The current $v_2$ measurement can not distinguish the deviations
of elliptic flow $\eta^{F(B)}$ from jet  effect of $\xi$ in
experiment, since they are mixed together. One can redefine
total $V_2(p_T)$ in forward ($F$) and backward ($B$) rapidity
bins,
       $V_2^F = (1 + \eta_F^{all}) \langle V_2^F\rangle$ , and
      $V_2^B = (1 + \eta_B^{all}) \langle V_2^B\rangle $,
  where $\eta_{F(B)}^{all}$ includes the fluctuation of elliptic flow and jet contribution. The F-B correlation can be rewritten as:
                \begin{align}
      & C_{FB} 
      = \frac{1 + \langle \eta_F^{all}\rangle  + \langle \eta_B^{all}\rangle  + \langle \eta_F^{all} \eta_B^{all}\rangle } {1 + \langle \eta_F^{all}\rangle  + \langle \eta_B^{all}\rangle  + \langle \eta_F^{all}\rangle
      \langle \eta_B^{all}\rangle }.
\label{eq15}
                   \end{align}
 With the assumption $\langle \eta_F^{all}\rangle  \,= \,\langle \eta_B^{all}\rangle  \,=0$, we can finally
 get the same expression as Eq.~(\ref{eq10}).

\begin{figure}[htbp]
\resizebox{8.5cm}{!}{\includegraphics{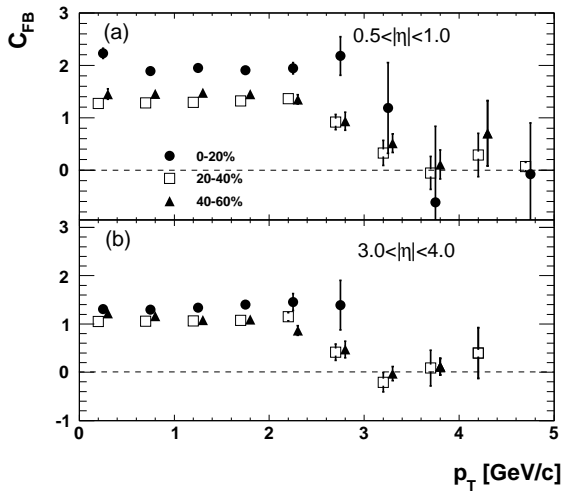}} \vspace{-0.4cm}
\caption{$C_{FB}$ for  200 GeV/c Au + Au collisions at 0-20$\%$
(circles), 20-40$\%$ (squares), and 40-60$\%$ (triangle)
centrality bins.
 (a)  0.5 $< |\eta| <$ 1.0, and (b)
 3.0$<|\eta|<$4.0 (Some symbols are slightly shifted in
$p_T$-axis for clarity).} \label{CFB_cen_v2_figure}
\vspace{-0.5cm}
\end{figure}

$C_{FB}$ has been simulated by using AMPT model for Au+Au at
$\sqrt{s_{NN}}$ = 200 GeV at the centrality bin 0-20$\%$. The
results are shown in Fig.~\ref{CFB_cen_v2_figure}, where different
symbols represent different  forward and backward pseudo-rapidity
bins. For example, $0.5<|\eta|<1.0$ represents that the forward
bin is within $0.5<\eta< 1.0$ and while the backward bin is within
$-1.0<\eta<-0.5$. $C_{FB}$ remains constant for different rapidity
gaps at low $p_T$ ($p_T < 2$ GeV/$c$) in 0-20$\%$ centrality bin.
To our surprise, $C_{FB} $ appears to be remarkably different from
the prediction of Ref.~\cite{Liao}, nearly two times as LK's
(unity) at low $p_T$ ($p_T < 2$ GeV/$c$) for the mid-rapidity gap,
and decreases with rapidity gap. The reason, as shown in Eq.
(\ref{eq6}) and (\ref{eq10}), is that the correlation of the
fluctuation of F-B elliptic flow $\langle \eta^F \eta^B\rangle $
is nonzero, i.e. $\langle \eta^F \eta^B\rangle $ is close to
$\frac{1}{4}$ at low $p_T$ for mid-rapidity gap at 0-20$\%$
centrality bin. As rapidity gap rises to 7.0
(Fig.~\ref{CFB_cen_v2_figure}(b)), $C_{FB} $ falls to 1.2, which
indicates $\langle \eta^F \eta^B\rangle $ becomes weak for a large
rapidity gap.

Fig.~\ref{CFB_cen_v2_figure} also displays  that $C_{FB}$ is much
larger for central collisions than non-central collisions at low
$p_T$, which indicates that  F-B elliptic flow fluctuation is
strongly correlated in the most central collisions. This trend is
consistent with  centrality dependence of elliptic flow
fluctuation ~\cite{centrality_dependence}. In addition, $\langle
\eta^F \eta^B\rangle $ is near to a positive constant at low $p_T$
for the given rapidity bins and centrality. However, at high $p_T$
($p_T > 4$ GeV/$c$), $C_{FB} $ is close to zero, indicating that
only jet contributes to the observable $v_2$ and $\langle \eta^F
\eta^B\rangle $ is zero. The results are similar with LK's finding
here. Therefore, $\langle \eta^F \eta^B\rangle $ should positively
decreases in intermediate $p_T$ region ($2 < p_T < 4$ GeV/$c$). It
means that $C_{FB} $ can not be used to directly measure the jet
contribution to observed $v_2$ without the knowledge of $\langle
\eta^F \eta^B\rangle $, especially for mid-rapidity gap.

The same tend is also seen in Fig.~\ref{CFB_cen_v2_figure}(b),
rapidity bin $3.0 < |\eta| < 4.0$. For centrality bins 20-40$\%$
and 40-60$\%$, $C_{FB} $ looks near to unity at low $p_T$, which
means $\langle \eta^F \eta^B\rangle $ is close to zero. The
structures are similar with LK's finding here. Therefore, $C_{FB}
$ with large rapidity gap is a clean probe to extract jet
contribution to elliptic flow, as Ref.\cite{Liao} expected.

\begin{figure}[htbp]
\resizebox{8cm}{!}{\includegraphics{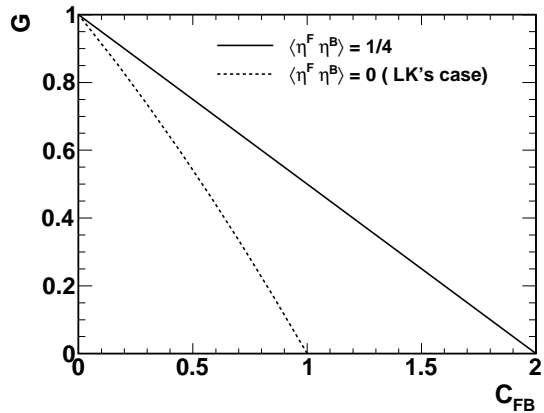}} \vspace{-0.4cm}
\caption{The relation of $G$ and $C_{FB}$  for different F-B
elliptic flow fluctuation correlation $\langle \eta^F
\eta^B\rangle $ in 200GeV/c Au + Au collisions, where $\langle
\eta^F \eta^B\rangle $ = $\frac{1}{4}$ corresponds to the case
with rapidity gap 0.5 $<  |\eta| <$ 1.0 at centrality bin
0-20$\%$. } \label{CFB_G_figure} \vspace{0.5cm}
\end{figure}

As we discussed in previous section,  two limits of $C_{FB}$, i.e.
$C_{FB} = 1 - G^2$ and $C_{FB} = 2(1 - G)$ which  can be deduced
assuming
 $\langle \eta^F \eta^B\rangle $ approaching to zero at high $p_T$
or $\langle \eta^F \eta^B\rangle $ approaching to $\frac{1}{4}$ at
low $p_T$, respectively.  Fig.~\ref{CFB_G_figure} shows the above
two limits. We can see that the factor $G$, which reflects the jet
contribution to the observed $v_2$, cannot be strictly quantified
by the observed $C_{FB}$. This gives  only a range of $G$-value
which  can be estimated for mid-rapidity gap. This is due to the
reason that  we do not know the quantitative relation of $\langle
\eta^F \eta^B\rangle $ and $G$ for intermediate $p_T$ region. With
the decreasing of $C_{FB}$, the uncertain range of $G$ becomes
narrower.

\begin{figure}[htbp]
\resizebox{8.8cm}{!}{\includegraphics{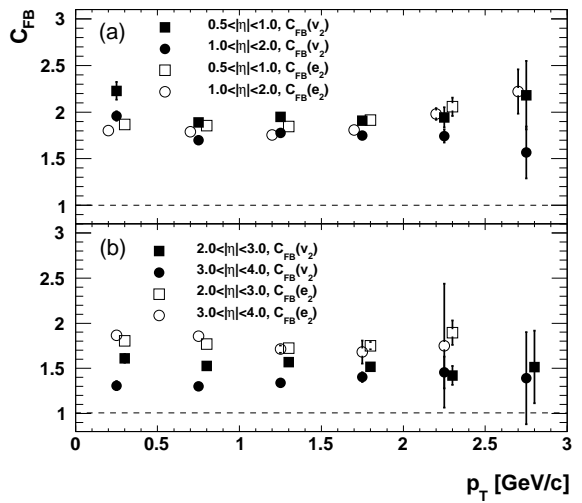}} \vspace{-0.4cm}
\caption{$C_{FB} $ for initial eccentricity (open symbols) and
elliptic flow (solid symbols)  in Au + Au collisions at
$\sqrt{s_{NN}}$ = 200 GeV at centrality bin 0-20$\%$ within
different rapidity gaps represented by different symbols. Some
symbols are slightly shifted in $p_T$-axis for clarity). }
\label{CFB_cen020_e2_v2_figure} \vspace{0.5cm}
\end{figure}

On the other hand, elliptic flow, as is shown in
Ref.~\cite{Voloshin,eccentricity_v2}, is converted from initial
collision geometry. It means that fluctuation of elliptic flow
should reflect the information on fluctuation in the initial state
geometry. Moreover, the fluctuation due to the magnitude and
centrality dependence of observed elliptic flow  are consistent
with fluctuation of the initial geometry shape of the collision
region~\cite{AlverPRC81}. It has been found that such a initial
geometry irregularity can be transferred into final momentum
anisotropies  by strong partonic interactions~\cite{ampthotspots,
triangularflow} (eg.  triangular flow). It becomes more
interesting to compare $C_{FB} $ for elliptic flow ($v_2$) and
initial eccentricity ($e_2$).

In AMPT model, partons, which are melted from string, are
initially distributed in momentum  and coordinate spaces. We can
similarly define $C_{FB} $ for initial eccentricity ($e_2$) as
     $ C_{FB} (e_2) = \frac{\langle E_2^F E_2^B\rangle }{\langle E_2^F\rangle  \langle E_2^B\rangle }$,
where $E_2^{F(B)}$ are the total eccentricity ($E_2$) in forward
($F$) and backward ($B$) rapidity bins, and
     $ \langle E_2(p_T)\rangle  \,= \frac{\langle y^2-x^2\rangle }
      { \langle y^2+x^2\rangle }\langle \frac{dN}{p_{t}dp_t}\rangle
      $.

Similar to $V_2^{F(B)}$, $E_2^{F(B)}$ can be written as
    $  E_2^F = (\frac{1}{2}+\eta_F^{'})\langle E_2\rangle $ and
     $ E_2^B = (\frac{1}{2}+\eta_B^{'})\langle E_2\rangle $,
where $\eta_{F(B)}^{'}$ represents random deviations from the
total eccentricity in forward (backward) rapidity bin. If one
compare with the former analysis of $v_2$,  $C_{FB}(e_2)$ presents
similar trend to $C_{FB}(v_2)$, if the correlation between
fluctuation of forward and backward elliptic flow is mainly due to
the initial correlation between fluctuation of forward and
backward eccentricity.

As shown in Fig.~\ref{CFB_cen020_e2_v2_figure}(a), it is an
obvious fact that $C_{FB}(v_2)$ is efficiently transferred from
$C_{FB}(e_2)$ at small rapidity gap. However, at large rapidity
gap (in Fig.~\ref{CFB_cen020_e2_v2_figure}(b)), $C_{FB} (v_2)$ is
suppressed much deeply in contrast to $C_{FB}(e_2)$. This
indicates that parton cascade are not strong enough for larger
rapidity gap to transfer the F-B correlation of initial $e_2$
fluctuation  to final F-B correlation of $v_2$ fluctuation. This
is because of the fact that parton interactions are weak at higher
rapidity. Therefore, the measurement of $C_{FB}$ for elliptic flow
($v_2$) may give us more information about the correlation of
initial geometry fluctuation.

In conclusion, jet and flow fluctuation contribution to elliptic
flow are investigated by forward-backward elliptic anisotropy
correlation ($C_{FB} $) within the framework of a multi-phase
transport model. We found that the $C_{FB} $ correlation is
remarkably different from LK's and is about nearly double of that
proposed by Liao and Koch. It stems from the correlation between
fluctuation of forward and backward elliptic flow at low
transverse momentum, which is mainly due to the initial
correlation between fluctuation of forward and backward
eccentricity. This leads to an amendment of the $C_{FB}$ by a term
related to the correlation between the fluctuation of forward and
backward elliptic flow. The present study shows that $C_{FB}$
correlation decreases with rapidity gap and a suitable rapidity
gap for $C_{FB}$ correlation studies should be around $\pm$ 3.5.

This work was supported in part by  the NSFC of China under Grant
No. 11035009, 10979074, 10875159, and the Shanghai Development
Foundation for Science and Technology under contract No.
09JC1416800, and the Knowledge Innovation Project of the Chinese
Academy of Sciences under Grant No. KJCX2-EW-N01.


\end{CJK*}
\end{document}